\documentclass{arxiv}
\usepackage[table]{xcolor}
\usepackage{hyperref}
\newcommand {\myvec}[1] {{\mbox{\boldmath $#1$}}}
\DeclareMathOperator{\erf}{erf}
\interspeechcameraready 

\title{Sparse Binarization for Fast Keyword Spotting}

\name[affiliation={1}]{Jonathan}{Svirsky}
\name[affiliation={1}]{Uri}{Shaham}
\name[affiliation={1}]{Ofir}{Lindenbaum}

\address{
  $^1$Bar Ilan University, Israel} %\\
\email{ \{svirskj,ofir.lindenbaum,uri.shaham\}@biu.ac.il}


\begin{document}

\maketitle

\begin{abstract}
With the increasing prevalence of voice-activated devices and applications, keyword spotting (KWS) models enable users to interact with technology hands-free, enhancing convenience and accessibility in various contexts. Deploying KWS models on edge devices, such as smartphones and embedded systems, offers significant benefits for real-time applications, privacy, and bandwidth efficiency. However, these devices often possess limited computational power and memory. This necessitates optimizing neural network models for efficiency without significantly compromising their accuracy. To address these challenges, we propose a novel keyword-spotting model based on sparse input representation followed by a linear classifier. The model is four times faster than the previous state-of-the-art edge device-compatible model with better accuracy. We show that our method is also more robust in noisy environments while being fast. Our code is available at: \textcolor{green}{\href{https://github.com/jsvir/sparknet}{https://github.com/jsvir/sparknet}}.

\end{abstract}

\section{Introduction}

Keyword spotting (KWS), also known as wake word detection, plays a crucial role in enabling voice-activated applications on micro-controllers and edge devices. 
These compact models are designed to continuously listen for specific trigger words or phrases like "Hey Siri," "Okay Google," or custom commands in constrained environments. Given the limited computational resources, memory, and power available on such devices, KWS models are optimized for efficiency and low latency without compromising accuracy. Edge devices are designed to be always operational, constantly listening for specific trigger phrases. Upon detecting these wake-up words, the system can activate a more sophisticated speech recognition model, either locally on the device or remotely in the cloud. This workflow, inherent to edge computing, offers several advantages, including reduced power consumption, minimized bandwidth usage, enhanced privacy, improved reliability, cost efficiency, and decreased response times \cite{abadade2023comprehensive}. 

The optimization of KWS models involves leveraging lightweight neural network architectures, such as small-footprint Convolutional Neural Networks (CNNs) or Recurrent Neural Networks (RNNs), and techniques like quantization and pruning to reduce model size and computational demands. To be able to run KWS models on microprocessors, e.g., STM32, two main requirements should be fulfilled. The first one is the limited memory footprint: the random-access memory
(RAM) is used to store input/output, weights, and activation
data while running. The size of on-chip RAM (SRAM) varies
from 20 KB to 512 KB. The second one is the limited computing speed: the number of operations per second is limited. The CPU (Cortex-M) frequency is typically between 72 MHz and 216 MHz. These hardware constraints limit the neural network models in two ways: the parameter count and the number of operations.
To address overcome this limitation, small-footprint deep neural networks were proposed, e.g., \cite{majumdar2020matchboxnet, kim21l_interspeech, tang2018deep, wong2020tinyspeech, xu2020depthwise,rozner2023efficient}.

In this work, we leverage the \textbf{spar}se binarized representation learning for the \textbf{K}WS model. We design a fast neural network, \textbf{SparkNet}, that yields accurate results that are on par with those produced by state-of-the-art (SOTA) small models but with much fewer multiplications. While these models utilize the two-dimensional convolutions \cite{kim21l_interspeech} or attention blocks \cite{wong2020tinyspeech} to achieve accurate predictions, we get SOTA results with reduced computation burn by learning binary sparse representation based on one-dimensional convolutions. Intuitively, we propose to learn a dynamic binarization model that discards the non-informative features in the signal and preserves those that are helpful for prediction. Inspired by the recently proposed SG-VAD model \cite{svirsky2023sg} for voice activity detection task, we learn a sample-wise binary representation with a tiny neural network. However, instead of jointly training the auxiliary classifier that is fed with the binary representation multiplied by input samples, our classifier is a single linear projection layer onto the target space and is fed by the predicted binary representation. Learning such a binarization of the data could be viewed as a hard compression of the input signal into a sparse representation with limited values in the $[0,1]$ range. In addition, we test the model robustness at different SNR scenarios, analyze its performance with a further decrease in the number of parameters, and provide an ablation study on model components. In the following sections, we present the model in detail and the experimental results for applying the method to the commonly used KWS benchmarks.

\begin{figure}[t]
% \begin{minipage}{.9\linewidth}
  \centering
    \includegraphics[width=0.85\columnwidth]{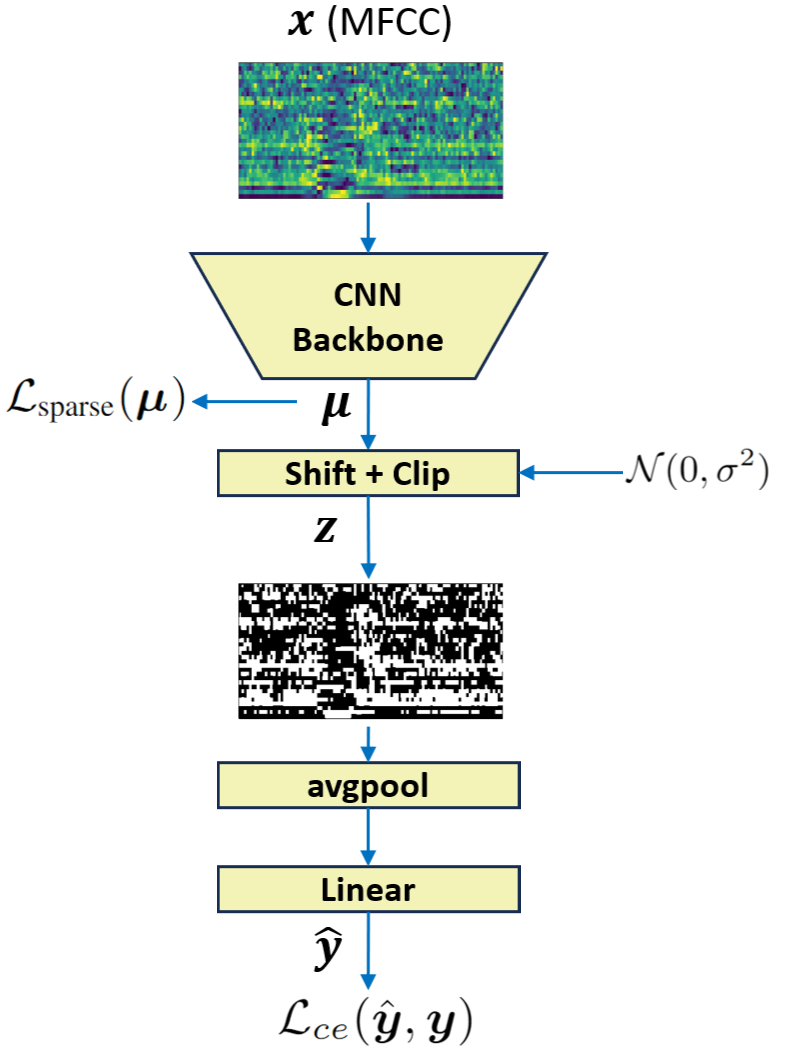}    
    \caption{\textit{The proposed SparkNet model learns a binary representation $\myvec{z}$ by using reparameterization trick. The random noise is added to the predicted $\myvec{\mu} \in [-1,1]^{F \times T}$ and approximate binary $\myvec{z}$ is obtained by centering the values on $0.5$ and clipping
    n the interval $[0, 1]$. 
    Only most informative features for classification task get positive values in $\myvec{z}$.}}
    \label{fig:train}
     \vskip -0.2 in
\end{figure}

\section{Proposed Method}
\subsection{Method Overview}

In a recent study \cite{svirsky2023sg}, the authors introduced a network that uses binary masking to reduce noise in an input signal for a classification task. They also demonstrated that the network's learned \textit{gates} can detect voice activity. Inspired by this idea, we developed a model that learns to binarize the input time-frequency representation to enable efficient and accurate Keyword Spotting (KWS). The model is optimized to predict local gates that depend on the input samples. These gates open for informative features in the input representation space and close for noisy or irrelevant features. Surprisingly, the binary gates are sufficient for predicting keywords with high accuracy, even with a single linear classification layer.

The proposed model utilizes a small neural network as an input binarizer that comes equipped with a single linear projection layer. This layer allows the network to project the binarized representation into the target categories space. Unlike other dynamic sparsification methods such as \cite{yang2022locally} and \cite{svirsky2023sg}, our model predicts the target without the need for element-wise multiplication of input samples and gates during training. Instead, it relies on the learned approximate binary representation. This reduces the computational complexity while preserving accuracy. 

The local binarizing module predicts binary $\myvec{z}=\myvec{z}(\myvec{x})$ of the same size as its input $\myvec{x}$ while the linear classifier predicts $\hat{y}$ based on average pooling of predicted $\myvec{z}$ in the time dimension. The model's design is based on the results we present in the ablation study, where we found that $\myvec{z}$ themselves conditioned on the input samples $\myvec{x}$ are sufficient for the classifier to produce accurate results. This observation is consistent with the inference architecture of the SG-VAD model, where only the binary mask serves as a classifier for the speech detection task.

\subsection{SparkNet Architecture}

The architecture of our model, SparkNet, is presented in Table \ref{tab:sgnet}. The main building block of the model is a time channel separable convolution \cite{kriman2020quartznet}, an implementation of the depth-wise separable convolutions. One-dimensional (1D) time channel separable convolutions can be separated into two layers. The first layer is a 1D depth-wise convolutional layer with a kernel length of $K$. This layer operates on each channel individually across $K$ time frames. The second layer is a point-wise convolutional layer that operates on each time frame independently but across all channels. By using these two layers, the model can process data in time-frequency format ($T \times F$). This completely separates the time and channel-wise parts of the convolution.  

The model includes 4 blocks with a 1D time-channel separable convolution layer followed by batch normalization and rectified non-linearity. The last three blocks also include the residual connections. To increase the model effective size of receptive fields, the convolutions are equipped with kernels of widths $K \in \{ 11, 15, 19, 29\}$. In addition to the first 4 blocks, a 1$\times$1 convolution layer is used as an output layer, followed by batch normalization and tanh activation.

The proposed model accepts an input sample constructed from Mel-frequency cepstral coefficients (MFCC) features matrix  $\myvec{x}_i \in \mathbb{R}^{F \times T}$, where $F$ is the number of frequency bins, and $T$ is the number of overlapping time frames.
The output linear layer projects the averaged representations across time to 12 keyword categories, where 10 are target words, one is an "unknown" category, and one is a "silence" category.

\subsection{Sparse Binarized Representation Learning}
\label{sec:sgnet}

The convolutional neural network backbone learns a representation $\myvec{\mu}_i \in [-1,1]^{F \times T}$ which is then converted to the approximate Bernoulli variables $\myvec{z}_i$, or \textit{binarized} version of $\myvec{x}$. To achieve that, we utilize a Gaussian-based relaxation of Bernoulli variables \cite{yamada2020feature}. The relaxation relies on the reparameterization trick \cite{miller2017reducing, figurnov2018implicit} and was demonstrated effective in several applications \cite{jana2023support,svirsky2023interpretable,lindenbaum2021l0}. During the training, the conversion is done by adding random noise matrix $\myvec{\epsilon}_i^{F \times T}$ to the shifted $\myvec{x}_i$ and clipping the values by range $[0,1]$:
\begin{equation}\label{eq:gates}
 \myvec{z}_i = \max(0, \min(1, 0.5 + \myvec{\mu}_i + \myvec{\epsilon}_i)),
\end{equation}
where each value in the matrix $\myvec{\epsilon}_i$ is drawn from $\mathcal{N}(0, \sigma^2)$ and $\sigma=0.5$ is fixed throughout training. To encourage the model to produce sparse $\myvec{z}_i$ matrix, it is trained with the regularization loss term $\mathcal{L}_{\text{sparse}}(\myvec{z}_i)=||\myvec{z}_i||_0$. The loss term is approximated using a double sum in terms of the Gaussian error function ($\erf$) on values of $\myvec{\mu}_i$: 
\begin{equation*}
    {\cal{L}}_{\text{sparse}} (\myvec{\mu}_i) = \frac{1}{F \times T}\sum^F_{f=1} \sum^T_{t=1}\left(\frac{1}{2} - \frac{1}{2} \erf\left(-\frac{\mu_{f,t} + 0.5}{\sqrt{2}\sigma}\right) \right).
\label{eq:reg}
\end{equation*}

When using the $\cal{L}_{\text{sparse}}$ term, the model tries to sparsify time-frequency bins that are associated with nuisance input features. This helps to simplify the data and make the model more efficient. Additionally, the classification layer propagates gradients to the binarization module for features that are necessary to accurately predict the targets.

\begin{table}[h]
  \caption{SparkNet layers with input shape of frequency$\times$time, ($F \times T$), the number of output channels $c$, stride $s$ and dilation $d$. The model consists of 4 blocks with time-channel separable (TCS) convolutions followed by batch normalization and ReLU activation, and an additional 1$\times$1 convolution with batch normalization, tanh activation, and output linear layer. For the base model, we use $C=16$.}
  \label{tab:sgnet}
  \centering
  \resizebox{.85\linewidth}{!}{
  \begin{tabular}{|c l c c c c|}
    \toprule
    \textbf{Input} & \textbf{Layer} & c & s & d & residual\\
    \midrule
    F$\times$T & Conv1D TCS (K=11)-BN-ReLU  & C & 1 & 1 & $\times$\\
    C$\times$T & Conv1D TCS (K=15)-BN-ReLU  & C & 1 & 1 & v\\
    C$\times$T & Conv1D TCS (K=19)-BN-ReLU  & C & 1 & 1 & v\\
    C$\times$T & Conv1D TCS (K=29)-BN-ReLU  & C & 1 & 1 & v\\
    C$\times$T & Conv1D (K=1)-BN-Tanh & F & 1 & 1 & $\times$\\
    F$\times$T & avgpool  & F & - & - & -\\
    F$\times$1 & Linear  & 12 & - & - & - \\
    \bottomrule
  \end{tabular}}
  
\end{table}

\subsection{Classification Learning}

We propose to train a linear classifier based on the sparse binary representation learned by a neural network. We apply average pooling in the time dimension and project the obtained representation vector onto the target space by a single linear layer. The binarizer learns class-wise patterns in the original $F \times T$ dimension that correlate with target labels. Additionally, in our ablation study presented in Sec. \ref{sec:ablation}, we found that the auxiliary classifier or decoder training on binary mask multiplied by the original input sample is redundant in the KWS task.

\begin{equation}
\label{eq:loss}
    \mathcal{L} =\mathcal{L}_{\text{sparse}}(\myvec{z}) + \lambda \mathcal{L}_{\text{ce}}(\hat{\myvec{y}}, \myvec{y}) 
\end{equation}

To conclude, we train the model to minimize the loss in eq. (\ref{eq:loss}) where $\mathcal{L}_{\text{ce}}(\hat{\myvec{y}}, \myvec{y})$ is the cross-entropy loss between the model prediction $\hat{\myvec{y}}$ and target $\myvec{y}$. By learning the binary representation as a basis for the linear prediction layer, the model can be more robust to noisy environments because it focuses on the indicative time-frequency regions in MFCC representation. Table \ref{tab:snrs} shows that our model is more robust than the recently proposed BC-Resnet state-of-the-art model at different signal-to-noise ratios (SNRs), and we attribute this improvement to the sparse binarized representation learning. Additionally, in section \ref{sec:results}, we demonstrate that while being as accurate or slightly better, our model is much faster in terms of multiply-accumulate operations (MACs). This makes it more suitable to be run on very low resource edge devices.

\section{Experiments}
\subsection{Experimental Setup}

\hskip 0.1in \textbf{Datasets} \hskip 0.1in We evaluate our method on Google Speech Commands versions 1 (SC1) and 2 (SC2) datasets \cite{warden2018speech}. Version 1 contains 64,727 utterances from 1,881 speakers for 30 word categories and version 2 has 105,829 utterances from 2,618 speakers for 35 words. Each utterance is 1 sec long, and the sampling rate is 16 $kHz$. We divide the both datasets into training, validation and testing set based on the validation and testing file lists \cite{warden2018speech}. 
The datasets are pre-processed following \cite{kim21l_interspeech, tang2018deep,warden2018speech} for the keyword spotting task where only 10 words are interesting targets, specifically: “Yes”, “No”, “Up”, “Down”, “Left”, “Right”, “On”, “Off”, “Stop”, and “Go with two additional classes "Unknown" (words from remaining twenty categories our of 35) and "Silence" (silence or background noise). We follow the common settings of \cite{kim21l_interspeech, warden2018speech} and re-balance the “Unknown Word” and “Silence” with the average number of utterances in the remaining ten classes. The audio segments are processed by extracting Mel-frequency cepstral coefficients (MFCC) features with $F=32$ bins.

\textbf{Evaluation} \hskip 0.1in  We use the top-1 accuracy metric to evaluate the model efficacy and we compute multiply accumulate operations (MACs) to evaluate the efficiency of the method\footnote{Calculated by \text{https://github.com/Lyken17/pytorch-OpCounter/} }.

\textbf{Robustness to noise} \hskip 0.1in 
We evaluate the model's robustness to noisy environments. For that goal we train the model on the SC2 train set without background noise augmentation and prepare a noisy version of the SC2 test set by adding random background noise to the original clean test set obtained from Freesound \cite{font2013freesound} and consists of 35 categories of noises, in total about 2,100 variable-length audio segments, such as ”motorcycle,” ”Bus,” and ”Static”. For each sample in the test set, we randomly pick noise and add it with SNR value in [0, 5, 10, 15, 20] decibels. We repeat this process 10 times with different random seeds. We use the same noisy test set to evaluate the BC-Resnet-0.625 trained on the same data as our model.

\textbf{Training Setup} \hskip 0.1in
 The input was augmented with time shift perturbations in the range of T = $[-100; 100]$ $ms$ and white noise of magnitude $[-90;-46]$ $dB$ with a probability of $80\%$. In addition, we add random background noise for SparkNet[$C=32$] with a probability of $80\%$. The model was trained with the SGD optimizer with momentum = $0.9$ and weight decay = $1e-3$. We utilized the Warmup-Hold-Decay learning rate schedule \cite{he2019bag} with a warm-up ratio of 5\%, a hold ratio of 40\%, and a polynomial (2nd order) decay for the remaining 85\% of the schedule. A maximum learning rate of $1e-2$ and a minimum learning rate of $1e-6$ were used. We trained all models for 200 epochs on a single NVIDIA GeForce GTX 1080 Ti with a batch size of 128. We use the fixed value for $ \lambda=1e+2$ for all experiments. The implementation of our model is based on the NeMo toolkit \cite{kuchaiev2019nemo}.

\begin{table*}[h]
  \caption{We compare our model against BC-Resnet for robustness to noisy environments while being $\sim \times4$ faster. The noisy tests are generated at different SNRs and present the mean accuracy on each test and standard deviation on 10 test variants for each SNR. }
  \label{tab:snrs}
  \centering
  \resizebox{.70\linewidth}{!}{
  \begin{tabular}{|c c c c c c c |}
    \toprule
    \textbf{Model / SNR} & \textbf{0 db} & \textbf{5 db} & \textbf{10 db} & \textbf{15 db} & \textbf{20 db} & \textbf{clean} \\
    \midrule
    BC-ResNet-0.625& 75.72 $\pm$ 2.73 & 84.47 $\pm$ 1.82 & 90.68 $\pm$ 1.35 & 92.67 $\pm$ 0.45 & 94.18 $\pm$ 0.20 & 95.40 $\pm$ 0.31\\
    \rowcolor[HTML]{DDDDDD} SparkNet[$C=16$] & 75.98 $\pm$ 2.66 & 85.05 $\pm$ 1.76 & 91.37 $\pm$ 1.16 & 93.58 $\pm$ 0.33 & 94.63 $\pm$ 0.30 & 95.70 $\pm$ 0.17\\	
    \bottomrule
  \end{tabular}}
\end{table*}

\begin{figure*}[t!]
  \centering
  \includegraphics[trim={20cm 10cm 20cm 54cm},clip, width=0.99\linewidth]{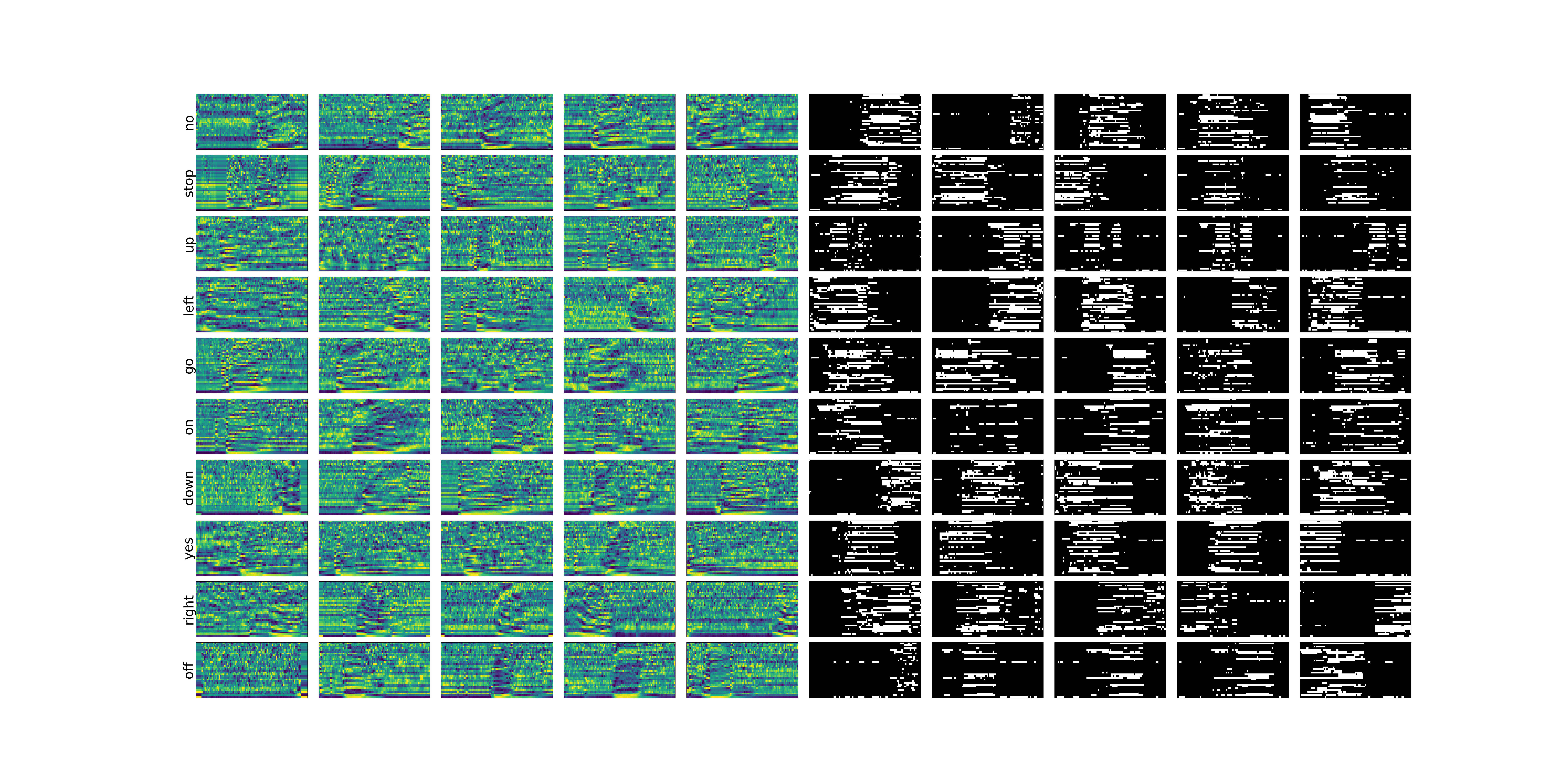}
  (a) \hskip 2.66in (b)
  \caption{ 5 randomly chosen words from 5 categories in MFCC representation (a) and the predicted binary representation (b).}
 \label{fig:masks}
   \vskip -0.1in
 \end{figure*}
 
\subsection{Results}
\label{sec:results}
In the first experiment, we compare two versions of our model with the SOTA low-resource BC-ResNet and other recently proposed methods. For this comparison, we train two version of our model, the first one with number of channels $C=16$ and in total $\sim4.6K$ parameters and the second one with $C=32$ and $\sim11K$ parameters. The former one is compared against BC-Resnet-0.625 which is a reduced version of BC-Resnet-1 following the definition in \cite{kim21l_interspeech}. The second small model is TinySpeech-Z \cite{wong2020tinyspeech} which has no public implementation and the results are borrowed from the published paper. For the second, extended version of SparkNet we present the results of BC-ResNet-1, DS-Resnet \cite{xu2020depthwise}, res8-narrow \cite{tang2018deep} and TinySpeech-X \cite{wong2020tinyspeech}. All models have up to $20K$ parameters. 
The accuracy and measured MAC results are provided in Table \ref{tab:results}. The base version of the proposed model with $C=16$ is $\times$4 faster than the corresponding BC-ResNet variant while producing a slightly improved accuracy. Our model with $C=32$ is $\times3$ faster than BC-ResNet-1 and $\times5$ - than DS-ResNet10. Additionally, the base version SparkNet[$C=16$] is more accurate than corresponding BC-ResNet version and $\times4$ faster than it in terms of MACs. 

In the second experiment we test our model and BC-ResNet with $\sim 4.6K$ parameters on noisy test set where noises are added with varying SNRs. The mean and std accuracy results over 10 versions for each SNR are presented in the Table \ref{tab:snrs}. It could be seen that our model is more robust to the noise than BC-ResNet. 

Finally, we challenge our model by reducing its size further by reducing the number of output filters down to $C=4$ in the intermediate layers. We train all versions on both SC1 and SC2 and test on the official test sets. The accuracy results are provided in Table \ref{tab:sgnet_results}. The tiny version of the model with only $\sim 1.4K$ parameters and $105K$ MACs is able to produce a mean accuracy of $~83\%$.

\begin{table}[h]
  \caption{Comparison of different methods trained on Google Speech Commands v2 dataset and tested on the official test set of keyword spotting task with 12 targets. While being with the same accuracy and number parameters our model has $\times$4 less multiply-accumulate operations (MACs)}
  \label{tab:results}
  \centering
  \resizebox{.99\linewidth}{!}{
  \begin{tabular}{|l | l l | c c |}
    \toprule
    & & & \multicolumn{2}{c|}{\textbf{Accuracy}} \\
    \textbf{Model} & \textbf{Params} &  \textbf{MACs}  & \textbf{SC1} & \textbf{SC2}\\
    \midrule
     \midrule
     TinySpeech-X \cite{wong2020tinyspeech} & 10.8K & 10.9M & 94.6 $\pm$ 0.00 & - \\
    res8-narrow \cite{tang2018deep} & 19.9K & 5.65M & 90.1 $\pm$ 0.98 & - \\
    DS-ResNet10 \cite{xu2020depthwise} & 10K & 5.8M & 95.2 $\pm$ 0.36 & - \\ 
    BC-ResNet-1 \cite{kim21l_interspeech}  & 9,232 & 3.6M & \textbf{96.6 }$\pm$ 0.21 & 96.9 $\pm$ 0.30 \\ 
    \rowcolor[HTML]{DDDDDD} SparkNet[$C=32$] & 11,500 & \textbf{1.2M} &  96.2 $\pm$ 0.19 & \textbf{97.0} $\pm$ 0.18  \\  %1170368.0
    \midrule
    TinySpeech-Z \cite{wong2020tinyspeech} & 2.7K & 2.6M & 92.4 $\pm$ 0.00  & - \\ 
    BC-ResNet-0.625 \cite{kim21l_interspeech} & 4,585 & 1.9M & 95.2 $\pm$ 0.37 & 95.4 $\pm$ 0.31 \\ 
    \rowcolor[HTML]{DDDDDD} SparkNet[$C=16$] & 4,636 & \textbf{454.5K} & \textbf{95.3 }$\pm$ 0.33 & \textbf{95.7} $\pm$ 0.17 \\ %454480
    \bottomrule
  \end{tabular}}
  \vskip -0.1in
\end{table}

\begin{table}[th]
  \caption{The accuracy of the proposed model as a function of number of parameters. The smallest model is still able to produce $\sim 83\%$ accuracy by using only 1.4K parameters and $105K$ MACs.}
  \label{tab:sgnet_results}
  \centering
  \resizebox{.85\linewidth}{!}{
  \begin{tabular}{|l | l l | c c |}
    \toprule
    & & & \multicolumn{2}{c|}{\textbf{Accuracy}} \\
    \textbf{Model Version} & \textbf{Params} &  \textbf{MACs}  & \textbf{SC1} & \textbf{SC2}\\
     \midrule
    SparkNet[$C=32$] & 11,500 & 1.2M &  96.2 $\pm$ 0.19 & 97.1 $\pm$ 0.30  \\  %1157440
    SparkNet[$C=16$] & 4,636 & 454.5K & 95.3 $\pm$ 0.33 & 95.7 $\pm$ 0.30 \\ %441,552
    SparkNet[$C=8$] & 2,292 & 190K & 91.6 $\pm$ 0.76 &  92.1 $\pm$ 0.33 \\ %190264
    SparkNet[$C=4$] & 1,416 & 105K & 82.3 $\pm$ 1.91  &  83.5 $\pm$ 0.60 \\ % 105020
    \bottomrule
  \end{tabular}}
\vskip -0.1in
\end{table}

\subsection{Ablation Study}
\label{sec:ablation}

To verify the essence of different parts of the proposed model we run experiments on the modified version of SparkNet on SC2 dataset. For all experiments we use SparkNet[$C=16$]. In the first experiment we check how an auxiliary larger classifier influences the training of the main model similarly to the SG-VAD model. We train SparkNet together with MatchboxNet-4x1x64 \cite{majumdar2020matchboxnet} which is fed by $\myvec{x} \odot \myvec{z}$ and predicts target labels as our main model. In this case, the loss becomes: $ \mathcal{L} =\mathcal{L}_{\text{sparse}}(\myvec{z}) + \lambda \mathcal{L}_{\text{ce}}(\hat{\myvec{y}}, \myvec{y}) +  \mathcal{L}_{\text{ce}}^{\text{aux}}(\hat{\myvec{y}}_{\text{aux}}, \myvec{y}) $ where $\hat{\myvec{y}}_{\text{aux}}$ are the prediction labels of MatchboxNet-4x1x64. Then we train our model without Bernoulli approximation by removing random noise adding to network output $\myvec{\mu}$, clipping and loss term $\mathcal{L}_{\text{sparse}}$. The model becomes a reduced version of MatchboxNet with less parameters. We also try to integrate a decoder auxiliary model that accepts $\myvec{x} \odot \myvec{z}$ which is trained for reconstruction task. The intuition, is that we would like to close the gates of correlating features and preserve the sufficient input signal for reconstruction task. So we add $\mathcal{L}_{\text{recon}}(\hat{\myvec{x}},\myvec{x})$ to the loss equation \ref{eq:loss} where $\hat{\myvec{x}}$ is the output from the decoder.

\begin{table}[th]
  \caption{Ablation study. Adding auxiliary classifier or decoder doesn't improve the model's accuracy. In addition, learning the binarized representation by Bernoulli approximation contributes to the model's performance.}
  \label{tab:example}
  \centering
  \resizebox{.65\linewidth}{!}{
  \begin{tabular}{|l c |}
    \toprule
    \textbf{Model Version} &  \textbf{Accuracy}\\
    \midrule
    SparkNet & \textbf{95.7} $\pm$ 0.17 \\
    SparkNet $+$ $\mathcal{L}_{\text{ce}}^{\text{aux}}(\hat{\myvec{y}}_{\text{aux}}, \myvec{y})$ & 95.6 $\pm$ 0.26 \\
    SparkNet $+$ $\mathcal{L}_{\text{recon}}(\hat{\myvec{x}},\myvec{x})$  & 95.5 $\pm$ 0.32  \\
        SparkNet w/o $\mathcal{L}_{\text{sparse}}$ & 94.7 $\pm$ 0.13  \\

    \bottomrule
  \end{tabular}}
  \vskip -0.1in
\end{table}

\subsection{Binarized Representation Qualitative Analysis}

We present in Figure \ref{fig:masks} the binarized samples for 5 samples from 5 word categories (up, right, no, stop, go, left). We extract MFCC features for each sample and add an additional layer of open gates (with value 1). As it could be seen, the model learns to localize the informative features and predicts sample-specific binary patterns.

\section{Conclusions}

We proposed a fast neural network for the KWS task with increased noise robustness. It's trained easily and makes accurate predictions while having a few number of parameters and an increased inference speed in terms of multiply accumulate operations. Our method is evaluated on two commonly used benchmarks for the KWS task and found to slightly outperform the previous SOTA small model in terms of accuracy measurements while being $\times 4$ faster in terms of MACs. We presented the improved robustness of the model to the noisy environments compared to the opponent model by testing the trained model at different SNRs. While our method is proposed for supervised KWS task, an interesting future extension for the model could be in self-supervised learning tasks where random masks as proposed in \cite{huang2022masked} are replaced by the learnable binary mask model.

\bibliographystyle{IEEEtran}
\bibliography{mybib}

% Generated by IEEEtran.bst, version: 1.13 (2008/09/30)
\begin{thebibliography}{10}
\providecommand{\url}[1]{#1}
\csname url@samestyle\endcsname
\providecommand{\newblock}{\relax}
\providecommand{\bibinfo}[2]{#2}
\providecommand{\BIBentrySTDinterwordspacing}{\spaceskip=0pt\relax}
\providecommand{\BIBentryALTinterwordstretchfactor}{4}
\providecommand{\BIBentryALTinterwordspacing}{\spaceskip=\fontdimen2\font plus
\BIBentryALTinterwordstretchfactor\fontdimen3\font minus \fontdimen4\font\relax}
\providecommand{\BIBforeignlanguage}[2]{{%
\expandafter\ifx\csname l@#1\endcsname\relax
\typeout{** WARNING: IEEEtran.bst: No hyphenation pattern has been}%
\typeout{** loaded for the language `#1'. Using the pattern for}%
\typeout{** the default language instead.}%
\else
\language=\csname l@#1\endcsname
\fi
#2}}
\providecommand{\BIBdecl}{\relax}
\BIBdecl

\bibitem{abadade2023comprehensive}
Y.~Abadade, A.~Temouden, H.~Bamoumen, N.~Benamar, Y.~Chtouki, and A.~S. Hafid, ``A comprehensive survey on tinyml,'' \emph{IEEE Access}, 2023.

\bibitem{majumdar2020matchboxnet}
S.~Majumdar and B.~Ginsburg, ``Matchboxnet: 1d time-channel separable convolutional neural network architecture for speech commands recognition,'' 2020.

\bibitem{kim21l_interspeech}
B.~Kim, S.~Chang, J.~Lee, and D.~Sung, ``{Broadcasted Residual Learning for Efficient Keyword Spotting},'' in \emph{Proc. Interspeech 2021}, 2021, pp. 4538--4542.

\bibitem{tang2018deep}
R.~Tang and J.~Lin, ``Deep residual learning for small-footprint keyword spotting,'' in \emph{2018 IEEE International Conference on Acoustics, Speech and Signal Processing (ICASSP)}.\hskip 1em plus 0.5em minus 0.4em\relax IEEE, 2018, pp. 5484--5488.

\bibitem{wong2020tinyspeech}
A.~Wong, M.~Famouri, M.~Pavlova, and S.~Surana, ``Tinyspeech: Attention condensers for deep speech recognition neural networks on edge devices,'' \emph{arXiv preprint arXiv:2008.04245}, 2020.

\bibitem{xu2020depthwise}
M.~Xu and X.-L. Zhang, ``Depthwise separable convolutional resnet with squeeze-and-excitation blocks for small-footprint keyword spotting,'' 2020.

\bibitem{rozner2023efficient}
A.~Rozner, B.~Battash, O.~Lindenbaum, and L.~Wolf, ``Efficient verification-based face identification,'' \emph{arXiv preprint arXiv:2312.13240}, 2023.

\bibitem{svirsky2023sg}
J.~Svirsky and O.~Lindenbaum, ``Sg-vad: Stochastic gates based speech activity detection,'' in \emph{ICASSP 2023-2023 IEEE International Conference on Acoustics, Speech and Signal Processing (ICASSP)}.\hskip 1em plus 0.5em minus 0.4em\relax IEEE, 2023, pp. 1--5.

\bibitem{yang2022locally}
J.~Yang, O.~Lindenbaum, and Y.~Kluger, ``Locally sparse neural networks for tabular biomedical data,'' in \emph{International Conference on Machine Learning}.\hskip 1em plus 0.5em minus 0.4em\relax PMLR, 2022, pp. 25\,123--25\,153.

\bibitem{kriman2020quartznet}
S.~Kriman, S.~Beliaev, B.~Ginsburg, J.~Huang, O.~Kuchaiev, V.~Lavrukhin, R.~Leary, J.~Li, and Y.~Zhang, ``Quartznet: Deep automatic speech recognition with 1d time-channel separable convolutions,'' in \emph{ICASSP 2020-2020 IEEE International Conference on Acoustics, Speech and Signal Processing (ICASSP)}.\hskip 1em plus 0.5em minus 0.4em\relax IEEE, 2020, pp. 6124--6128.

\bibitem{yamada2020feature}
Y.~Yamada, O.~Lindenbaum, S.~Negahban, and Y.~Kluger, ``Feature selection using stochastic gates,'' in \emph{International Conference on Machine Learning}.\hskip 1em plus 0.5em minus 0.4em\relax PMLR, 2020, pp. 10\,648--10\,659.

\bibitem{miller2017reducing}
A.~Miller, N.~Foti, A.~D'Amour, and R.~P. Adams, ``Reducing reparameterization gradient variance,'' \emph{Advances in Neural Information Processing Systems}, vol.~30, 2017.

\bibitem{figurnov2018implicit}
M.~Figurnov, S.~Mohamed, and A.~Mnih, ``Implicit reparameterization gradients,'' \emph{Advances in neural information processing systems}, vol.~31, 2018.

\bibitem{jana2023support}
S.~Jana, H.~Li, Y.~Yamada, and O.~Lindenbaum, ``Support recovery with projected stochastic gates: Theory and application for linear models,'' \emph{Signal Processing}, vol. 213, p. 109193, 2023.

\bibitem{svirsky2023interpretable}
J.~Svirsky and O.~Lindenbaum, ``Interpretable deep clustering,'' \emph{International Conference on Machine Learning}, 2024.

\bibitem{lindenbaum2021l0}
O.~Lindenbaum, M.~Salhov, A.~Averbuch, and Y.~Kluger, ``L0-sparse canonical correlation analysis,'' in \emph{International Conference on Learning Representations}, 2021.

\bibitem{warden2018speech}
P.~Warden, ``Speech commands: A dataset for limited-vocabulary speech recognition,'' \emph{arXiv preprint arXiv:1804.03209}, 2018.

\bibitem{font2013freesound}
F.~Font, G.~Roma, and X.~Serra, ``Freesound technical demo,'' in \emph{Proceedings of the 21st ACM international conference on Multimedia}, 2013, pp. 411--412.

\bibitem{he2019bag}
T.~He, Z.~Zhang, H.~Zhang, Z.~Zhang, J.~Xie, and M.~Li, ``Bag of tricks for image classification with convolutional neural networks,'' in \emph{Proceedings of the IEEE/CVF Conference on Computer Vision and Pattern Recognition}, 2019, pp. 558--567.

\bibitem{kuchaiev2019nemo}
O.~Kuchaiev, J.~Li, H.~Nguyen, O.~Hrinchuk, R.~Leary, B.~Ginsburg, S.~Kriman, S.~Beliaev, V.~Lavrukhin, J.~Cook \emph{et~al.}, ``Nemo: a toolkit for building ai applications using neural modules,'' \emph{arXiv preprint arXiv:1909.09577}, 2019.

\bibitem{huang2022masked}
P.-Y. Huang, H.~Xu, J.~Li, A.~Baevski, M.~Auli, W.~Galuba, F.~Metze, and C.~Feichtenhofer, ``Masked autoencoders that listen,'' \emph{arXiv preprint arXiv:2207.06405}, 2022.

\end{thebibliography}

\end{document}